\def\simlr{\mathbin{\;\raise1pt\hbox{$<$}\kern-8pt\lower3pt\hbox{$\sim$}\;}}
\def\simgr{\mathbin{\;\raise1pt\hbox{$>$}\kern-8pt\lower3pt\hbox{$\sim$}\;}}

\documentclass[a4paper]{jpconf}
\usepackage{graphicx}
\usepackage{ulem}       
\begin{document}
\title{Variable stars across the observational HR diagram}

\author{Laurent Eyer$^1$, Nami Mowlavi$^{1,2}$}

\address{$^1$ Observatoire de Gen\`eve, Universit\'e de Gen\`eve, 1290 Sauverny, Switzerland}
\address{$^2$ ISDC, Universit\'e de Gen\`eve, 1290 Versoix, Switzerland}

\ead{Laurent.Eyer@obs.unige.ch}

\begin{abstract}

An overview of pulsating variable stars across the observational Hertzprung-Russel (HR) diagram is presented, together with a summary of their global properties.
The HR diagram is presented with a third colour-coded dimension, visualizing the fraction of variable, the amplitude of variability or the period of variability.
The distribution of variable stars in the other observational diagrams, such as the Period-Amplitude diagram, is also presented.
Some of the progresses performed in the field of variable stars during the last decade are briefly summarized, and future projects that will improve our knowledge of variable stars are mentioned.

\end{abstract}

\section{Introduction}
There are in the literature several global descriptions of variable stars. We can mention four books, one by Richter, Wenzel and Hoffmeister (1985, out of print), one by Sterken and Jaschek \cite{sterkenetal1996}, one by Percy (\cite{percy2007}) and one, soon to come, by Aerts et al. \cite{aertsetal2007}. There have also been two excellent and detailed reviews by Gautschy and Saio (\cite{gautschyetal1995}, \cite{gautschyetal1996}) about a decade ago.

With the advent of Charge-Coupled Devices (CCDs) and large scale surveys like the Optical Gravitational Lensing Experiment (OGLE), the Massive Compact Halo Object project (MACHO) or the "Exp\'erience pour la Recherche d'Objets Sombres" (EROS), the subject of variable star studies is now under a remarkable expansion.
As we will see the expansion will continue in the coming decade because there are many future projects which will sample frequently large regions of the sky.
In other words there will be a larger number (hundreds of million to billion) of objects being measured repeatedly.
There is also a drastic improvement in the photometric quality with certain projects, exploring new regions of the variability levels.

With this expansion of observational facilities, new variable types have been discovered over the past 10 years, and important advances have been made in the understanding of pulsation theory.

In this review, we first present the different classes of variable stars in Sect.~\ref{Sect:tree}, and mention in Sect.~\ref{Sect:projects} the current and future projects that benefit to the field of variability research.
We then present several observational diagrams.
The classical Hertzprung-Russel (HR) diagram is tackled in Sect.~\ref{Sect:HR}, while other observational diagrams are addressed in Sect.~\ref{Sect:other diagrams}. 
A brief overview of the progresses recorded in the field during the last decade is presented in Sect.~\ref{Sect:progress}. A note on some points pertaining to stellar evolution models is given in Sect.~\ref{Sect:stellar evolution}. The paper ends with conclusions in Sect.~\ref{Sect:conclusions}.

\section{A variability tree}
\label{Sect:tree}

The \textsl{General Catalogue of Variable Stars} (GCVS) lists more than 100 types and subtypes of variable stars.
The variability properties of the light curves are very diverse, ranging from extremely regular pulsations like Cepheids to unique events such as cataclysmic phenomena or supernovae explosions, with times scales from the order of milliseconds for Gamma Ray Bursts to centuries for secular evolution, and with amplitude variations from parts per million for solar-like oscillations to many orders of magnitudes for the most energetic phenomena in the universe like hypernovae.

\begin{figure}
	\begin{center}
		\includegraphics[angle=90,width=170mm]{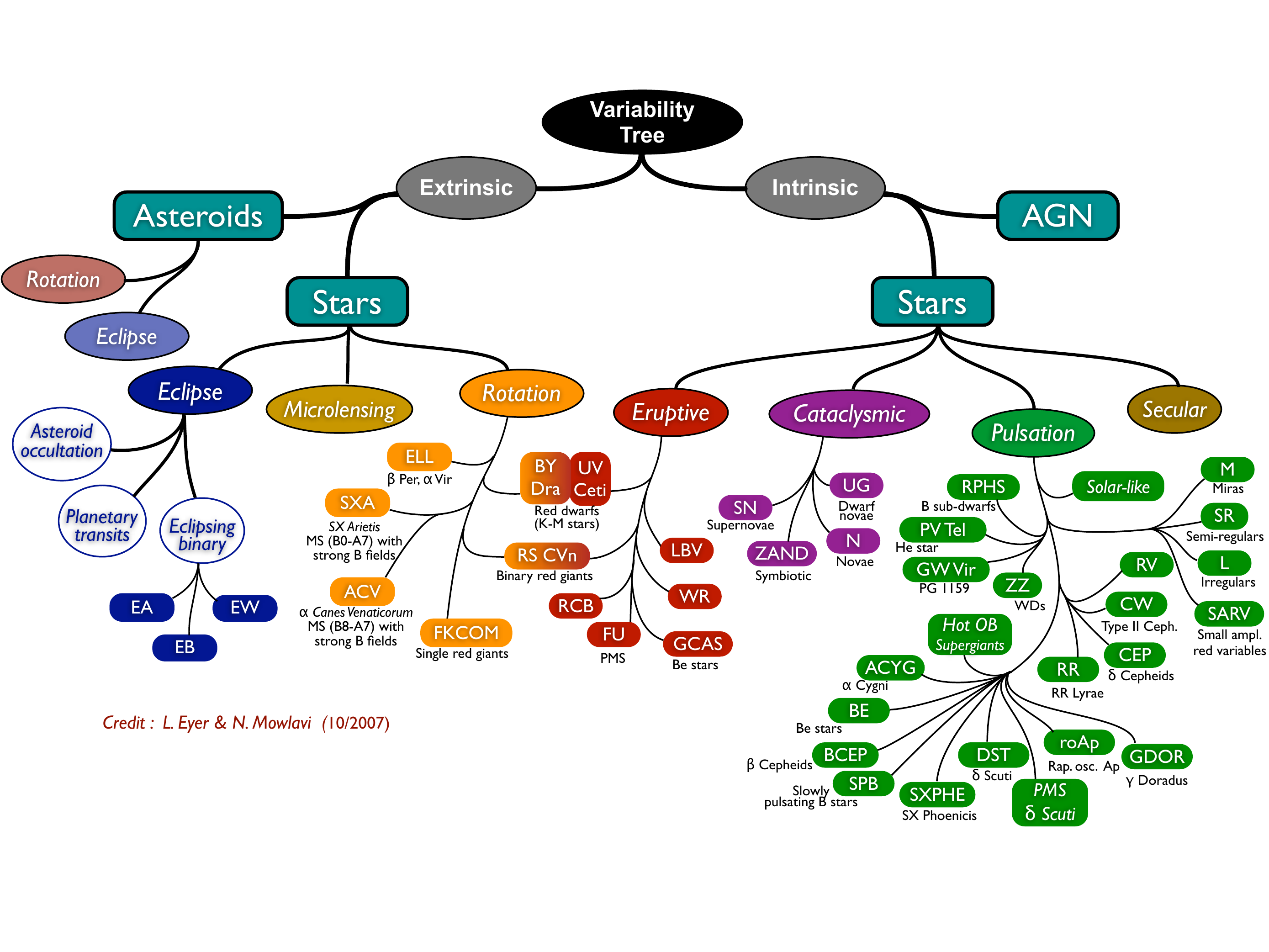}
	\end{center}
	\caption{\label{fig:vartree} Variability tree. A tentative organisation of variable objects.}
\end{figure}

Before covering the variable stars across the HR diagram, we attempt to group, in figure~\ref{fig:vartree}, the different types of variable sources that are present in the universe according to the physical phenomena at the origin of their variability.
Four division levels are introduced.
First, the "classical" division between {\it extrinsic} and {\it intrinsic} variables.
The former group identifies objects whose light curve variability results from a geometrical effect, either due to the position of the object relative to the observer or due to rotation.
The latter group, on the other hand, results from physical changes occurring in the object itself.


The second level of division concerns the type of object considered, being either asteroids, stars, or galaxies.
We should indeed not forget that most asteroids and Active Galactic Nuclei (AGN) are observed with variable light intensities.

The third level identifies the phenomenon at the origin of the variability.
Rotation, eclipses by a companion or by a foreground object, and microlensing effects are the phenomena considered for extrinsic variability, while intrinsic variable objects include eruptive stars, cataclysmic systems, stars displaying secular evolution and, last but not least, pulsating stars.

Finally, the fourth division groups objects of either similar physical nature or of similar photometric behaviour.
Two remarks must be given here. First, as in biology, the division into distinct classes is very useful, but has also its limits.
The transition between one class to another may be continuous in some cases (e.g. SARV, L, SR, M), leading to difficulties to attribute an object to one class or another if its properties overlap those of two contiguous classes of objects. Second, the importance of a good nomenclature must be stressed to avoid confusion.
It is of course difficult to find a good name for a newly discovered class of objects, as witnessed by the confusion found
in the literature about some classes of variable stars (some examples are shown in Sect.\ref{Sect:progress}).
It is probably time now to re-assess the conventional names given to the different known classes of variable objects.
This is certainly a difficult, though important, task for which a commission of the  IAU has been set up.
For the time being, we have mainly followed the nomenclature used in the {\it General Catalogue of Variable Stars} (GCVS).

This diagram is mostly oriented towards photometric variability. Obviously variable objects may have signatures in their spectra that are not necessarily detectable in their photometric light curve. Such is for example the case with Doppler shifts or line profile variations, which play an important role in astero-seismological studies or planet detection. Those cases are not reflected in figure~\ref{fig:vartree} and are not considered in this review.

Part of this conference is devoted more specifically to asteroseismology. Therefore we list the observational properties of the most common pulsating variable stars in table~\ref{tab:varstarchar}.

\begin{table}[h]
\caption{\label{tab:varstarchar} Most common pulsating variable stars and their fundamental characteristics. The periods are expressed in seconds (s), minutes (min), hours (h) or days (d).}
\begin{center}
\begin{tabular}{lllll}
\hline
Class name        & Period         & Amplitude (mag)    \\ 
\hline
$\alpha$ Cyg      & 1-50 d         &  few 0.1           \\ 
$\beta$ Cephei    & 2-12 h         &  up to     0.05    \\ 
Cepheids          & 2-70 d         &      0.1-1.5       \\ 
$\delta$ Scuti    & 30 min - 6 h   &  up to few 0.1     \\ 
EC 14026          & 80 s-8 min     &  up to    0.03     \\ 
$\gamma$ Dor      & 8 h -3 d       &  up to few 0.01    \\ 
GW Vir            & 5 min-80 min   &  up to 0.2         \\ 
PG1716 "Betsy"    & 30-150 min     &  up to $\sim$ 0.01 \\ 
roAp              & 5-20 min       &  up to     0.01    \\ 
RR Lyrae          & 5 h-1.1 d      &      0.2-2         \\ 
RV Tau            & 30-150  d      &  1-3               \\ 
SPB               & 0.5-5 d        &  up to     0.03    \\ 
SR, Mira          & 50-1000 d      &  up to 8            \\ 
V777 Her          & 2--16 min      &  up to 0.2         \\ 
W Vir             & 1-20 d         &    0.3-1.2         \\ 
ZZ Ceti           & 0.5--25 min    &  up to 0.2         \\ 
\end{tabular}
\end{center}
\end{table}

\section{Projects}
\label{Sect:projects}

Many large scale surveys have accumulated  multi-epoch photometric data during the last decade. 
Thanks to these surveys, the number of stars of different variability types is becoming large enough to allow a significant statistical description of their properties.
These surveys include ground based projects, such as the pioneering micro-lensing surveys MACHO or OGLE-I, II and III, or planetary transit surveys like the Hungarian-made Automated Telescope survey (HAT), the OGLE-III survey (which should soon become public), the Wide Angle Search for Planets survey  (SuperWasp), and the Trans-Atlantic Exploanet Survey (TreS).
Among the actions that can be done in order to benefit efficiently from those surveys are the release of both raw and reduced data into the public domain and the availability of a user-friendly interface that allows convenient extraction of the whole or part of the whole data set.
An example of data availability is provided by the All Sky Automated Survey (ASAS), which is currently continuing its data acquisition of the "bright" objects in the sky up to magnitude 14-15.

Efforts have also been developed during the last decade in multisite observations.
By obtaining as continuous observations as possible, aliasing effects are reduced and precise information on all frequencies of the variable objects can be obtained.
Projects such as the Delta Scuti Network project (DSN) and the Whole Earth Telescope project (WET) have taken this multi-site option.
Variable types that have mostly benefited from such programs include compact objects, $\delta$ Scuti stars and $\beta$ Cephei stars.
Obviously, those programs can only focus on a limited number of objects.

Several plans for future ground based facilities dedicated to large scale surveys will provide unprecedented material to improve our knowledge of stellar variability for a large number of objects.
Those include the Large Synoptic Survey Telescope (LSST, 8.4-meter, completion estimated for 2013) and the Panoramic Survey Telescope \& Rapid Response System (PanSTARRS, starting now a test phase and scientific studies lasting for 3.5 years). One project which deserves full attention is the Stellar Oscillations Network Group (SONG) multi-site project which will perform precise doppler-velocity observations for bright objects.
The Antarctica site has also been under focus and promises great potential.
There are many projects, such as the Seismic Interferometer Aiming to Measure Oscillations in the Interior of Stars (SIAMOIS), the International Concordia Explorer Telescope (ICE-T),  the Antarctica Search for Transiting Extrasolar Planets (ASTEP), and the International Robotic Antarctic Infrared Telescope (IRAIT) projects.

From space, the past missions such as the High Precision Parallax Collecting Satellite mission (Hipparcos; a survey type mission) or the Wide Field Infrared Explorer mission (WIRE; a pointing type mission with extremely precise photometry) have proved the advantages of operating above the atmosphere.
Reports on the on-going Microvariability and Oscillations of STars mission (MOST) have been shown during this meeting, as well as some remarkable preliminary results of the COnvection, ROtation and planetary Transits mission (CoRoT) that was launched at the end of last year (2006).
Despite the importance of those types of mission dedicated to the continuous monitoring of the sky with a high degree of photometric precision, it is surprising to see, retrospectively, how the realization of such missions may be uncertain on a time span of 8 years \cite{eyer2000}:
some missions that were foreseen to be cancelled may finally be launched (such as the NASA Kepler mission, now planed to be launched in February 2009), while other missions that were approved and seemed to lie on firm grounds were eventually cancelled (such as the ESA Eddington mission).

\begin{table}[h]
\caption{\label{tab:NumberOfVariables} 
         Number of variable stars observed by some projects.}
\begin{center}
\begin{tabular}{lrr}

\hline
\multicolumn{3}{l}{HIPPARCOS (All sky multi-epoch photometry)}\\
\hspace{3mm} {\it Main Mission}         & {\it 120,000}  &        \\  	
\hspace{3mm}	    Variables                   &  11,597	    & 9.66\% \\
\hspace{6mm}	  	Non-periodic/unsolved	&   5,542	& 4.62\% \\
\hspace{6mm}	  	Not investigated	  	 	&   3,343	& 2.79\% \\
\hspace{6mm}	  	Periodic	  	 			&   2,712	& 2.26\% \\
\hspace{9mm}	  	  	Delta Scuti	 	 	&	 186	  	& 0.16\% \\
\hspace{9mm}	  	  	RR Lyrae	 	 		&	 190	  	& 0.16\% \\
\hspace{9mm}	  	  	Cepheid	  			&	 273	  	& 0.23\% \\
\hspace{9mm}	  	  	Mira	 				&	 200	  	& 0.17\% \\
\hspace{9mm}	  	  	Other pulsating		&	 300	  	& 0.25\% \\
\hspace{9mm}	  	  	Eclisping Binaries	&	 917	  	& 0.76\% \\
\hspace{9mm}	  	  	Others	 			&	 646		& 0.54\% \\

\hline
\multicolumn{3}{l}{ASAS}\\
\hspace{3mm} {\it ASAS 1-2 ($I<13$)}     & {\it 140,000}  &        \\
\hspace{6mm}	    	Variables	  	  	    &       4,000	 & 2.86\% \\
\hspace{3mm}	 {\it  ASAS 3 ($V<15$)} as of mid-2005	  	& {\it 7,300,000} &  \\	  	
\hspace{6mm}	    	Variables	  	  	    &      24,936	  	 & 0.34\% \\
\hspace{9mm}	    	   Eclipsing Binaries	&       5,449	  	 & 0.07\% \\
\hspace{9mm}	    	   Mira					&       1,502	  	 & 0.02\% \\
\hspace{9mm}	    	   Regular pulsating	    &       2,416	  	 & 0.02\% \\
\hspace{9mm}	    	   Other (misc.)		    &       15,569	  	 & 0.21\% \\

\hline
\multicolumn{3}{l}{Tycho}\\
\hspace{3mm} {\it All objects}     & {\it 500,000}  &        \\
\hspace{6mm}     Variables         &     700       &   0.14\%     \\

\hline
\multicolumn{3}{l}{OGLE-I}\\
\hspace{3mm} {\it BULGE}     & {\it 2,000,000}  &        \\
\hspace{6mm}    $I_{\mathrm{mag}}<18$    &        400,000   &  \\	  	
\hspace{6mm}	    Variables	  	  	& 2,800	& 0.70\% \\
\hspace{9mm}	    Periodic	 			& 1,900 & 0.48\% \\
\hspace{9mm}	    Pulsating	 		&   250 & 0.06\% \\
\hspace{9mm}	    Misc.	 			&   900 &	 0.23\% \\
\hspace{9mm}	    Eclipsing Binaries	& 1,650 & 0.41\% \\
  
\hline
\multicolumn{3}{l}{OGLE-II}\\
\hspace{3mm} {\it BULGE}     & {\it 30,000,000}  &        \\
\hspace{6mm}    Variables    &        200,000   & 0.67\% \\
\hspace{3mm} {\it LMC/SMC}     & {\it 9,000,000}  &        \\
\hspace{6mm}    Variables    &        68,000   & 0.76\% \\
\hspace{9mm}	    RR Lyrae (SMC)	 	&		 571 & 0.006\% \\
\hspace{9mm}	    Cepheids	 			&   3'382	 & 0.034\% \\
\hspace{9mm}	    Eclipsing Binaries (SMC)	   & 1'459	& 0.016\% \\
\hspace{9mm}	    QSO (behind SMC)	 		  &	 $\sim$ 6	 & \\

\hline
\end{tabular}
\end{center}
\end{table}

We must note that some surveys that are designed for other fields of astronomy may nevertheless also contribute to the field of variability studies.
The Sloan Digital Sky Survey (SDSS), for example, whose primary goal was cosmology, allowed to perform variability studies on RR Lyrae stars \cite{ivezicetal2000}.
Some regions of the sky were covered by overlapping fields during the survey, leading to multi-epoch observations.
The SDSS also led to the discovery of asteroids through the detection of their relative motion in successive exposures \cite{ivezicetal2001}.
In addition, Blake et al. \cite{blakeetal2007} succeeded to find a very low mass eclipsing binary system in a time series containing only 30 epochs.

As an example of a future survey in space, we mention the Gaia mission, a cornerstone ESA mission with a foreseen launch date in 2011.
It will observe one billion stars, mostly in our Galaxy, and provide astrometric, photometric, spectrophotometric and radial velocity multi-epoch observations.
The requirements for the astrometric precision (parrallax error for a 5 year mission) mention "better than 24 micro arcsec for a G2 star at mag $V=15$".
The satellite will record a mean of about 80 measurements per object spread over 5 years.
The photometric precision is estimated to about 1-2 millimag per measurement for objects with a magnitude up to about $V=14$.
The satellite will have the opportunity to detect a large number of new variables, including short time pulsators of several tens of seconds.
If Gaia is operating at the precision level requested in the specifications, it will provide, among other results, HR diagrams with an unprecedented level of information.

The preparation of a mission is also an opportunity to develop new tools.
In the case of Gaia, for example, a model of the Galaxy is being developed that will provide, as a by-product, a realistic representation of the distribution of variable objects in the sky.
A better definition of the classes of variable objects than the one given in table~\ref{tab:varstarchar} must be established.
To this end, we are assembling data and defining probability distributions for the variability properties of different classes of variable objects in various observational diagrams (see Sects.~\ref{Sect:HR} and \ref{Sect:other diagrams}).
The goal is to predict the number and types of variables as well as their properties present in a given population.

The number of variables detected in a survey depends on the observed stellar population(s), on the covered range of magnitudes, on the length and the sampling law of the observation, on the overall instrument precision, and on the data processing methods used for variability analysis.
The number of variable objects detected in a few surveys is presented in table~\ref{tab:NumberOfVariables}. The largest fraction of variables is so far obtained by the Hipparcos mission, with 9.7\% of the 120,000 observed stars being variable.

\section{HR diagrams}
\label{Sect:HR}

\begin{figure}
	\begin{center}
		\includegraphics[width=120mm]{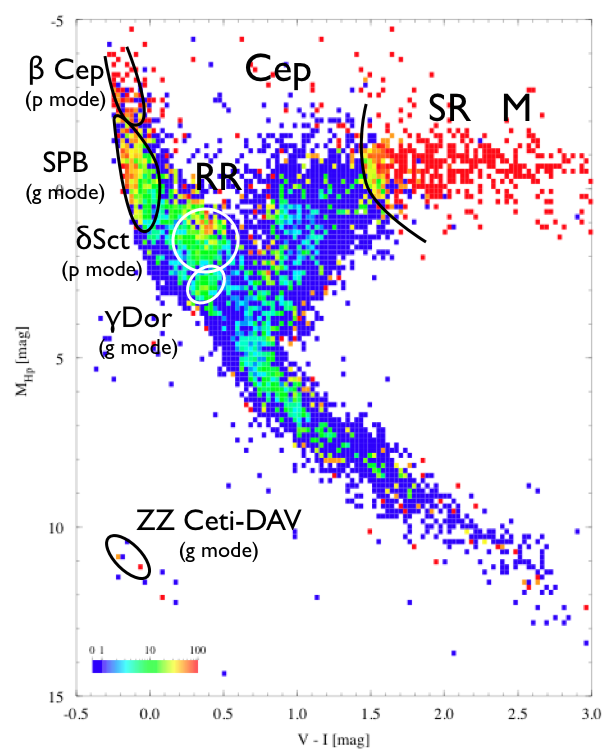}
	\end{center}
	\caption{\label{fig:HRfofv} Fraction of variable across the HR diagram -- from Hipparcos catalogue \cite{ESA1997}.}
\end{figure}

\begin{figure}
	\begin{center}
		\includegraphics[width=120mm]{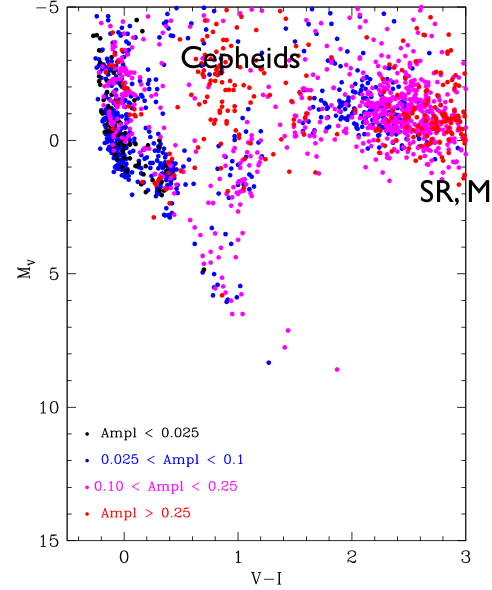}
	\end{center}
	\caption{\label{fig:HRamp} HR Diagram with colour coding for amplitudes. The data are taken from the Annexes "Unsolved variables" and "Periodic variables" in the Hipparcos catalogue.}
\end{figure}

\begin{figure}
	\begin{center}
		\includegraphics[width=120mm]{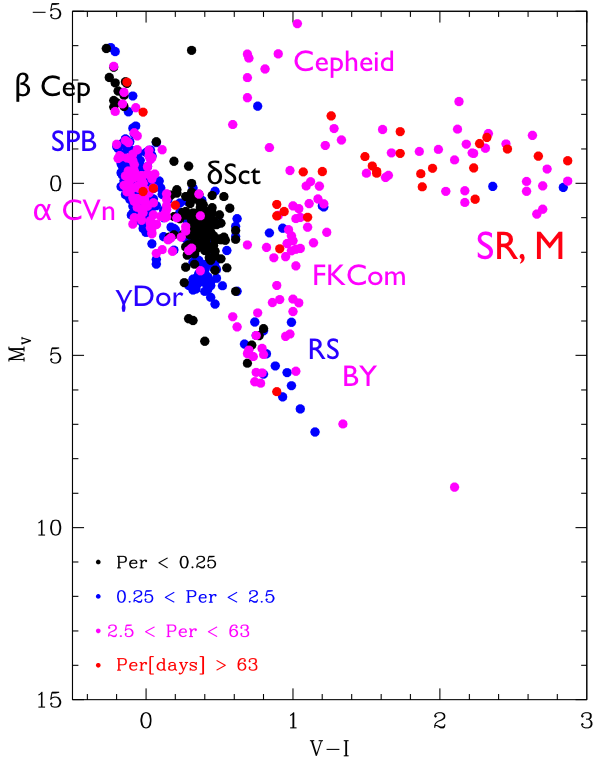}
	\end{center}
	\caption{\label{fig:HRper} HR Diagram with colour coding for periods. The data are taken from the Annex "Periodic variables" in the Hipparcos catalogue. Some types of pulsating and rotation induced variables are shown in the figure.}
\end{figure}

The HR diagram is often expressed in terms of the effective surface temperature and absolute bolometric luminosity, which are the physical quantities intrinsic of stars. The observed quantities, however, are usually the apparent magnitude and a colour index. There are several caveats when converting observable to intrinsic quantities.
The transformation of the apparent magnitude into an absolute magnitude requires the knowledge of the distance to the object, which can be obtained either with the parallax for objects close enough to the Earth, with some calibration procedure (Cepheids for example), or with the knowledge of the distance to the group of stars to which the object belongs, if any (this is the case for clusters, the magellanic Clouds, or Galaxies).
In addition, since the apparent magnitude is measured in a given band, a bolometric correction must be applied to infer the bolometric magnitude.

A most often used HR diagram is that of the absolute V magnitude $M_V$ versus the B-V colour. We must keep in mind that the morphology obtained in an HR diagram based on observable quantities varies - sometimes dramatically - depending on the spectral bands that are used for the colour and/or apparent luminosity. We must also keep in mind that some stars may display large variations in $M_V$, but tiny ones in the bolometric magnitude.

The interstellar extinction, metallicity, duplicity and rotational effects may also affect the morphology of the HR diagram. 
These effects can be far from being negligeable. The interpretation of an HR diagram may thus be delicate.


A third dimension can be added to the colour-magnitude HR diagram by using colours to scale a third parameter. For variable stars the third parameter can be the fraction of variable stars, the amplitude of variability, or its period. For Hipparcos data, such diagrams have been published in  Refs. \cite{ESA1997},  \cite{eyeretal1994} and  \cite{eyeretal1997}. The Hipparcos diagrams within these last two references were used with the survey part of Hipparcos catalogue (in order to avoid biases from the selected sample of Hipparcos) and excluding the eclipsing binaries. For the Geneva photometry database, they can be found in \cite{grenon1993}.  The diagrams presented in the Hipparcos catalogue (\cite{ESA1997}) were obtained using the absolute magnitude, while the other papers presented the diagrams in terms of the spectral and the luminosity classes. 

The distribution of the fraction of variable stars from the Hipparcos catalogue is shown in figure~\ref{fig:HRfofv}. That of the amplitudes and periods are displayed in figures~\ref{fig:HRamp} and \ref{fig:HRper}, respectively.


\begin{figure}
	\begin{center}
		\includegraphics[width=120mm]{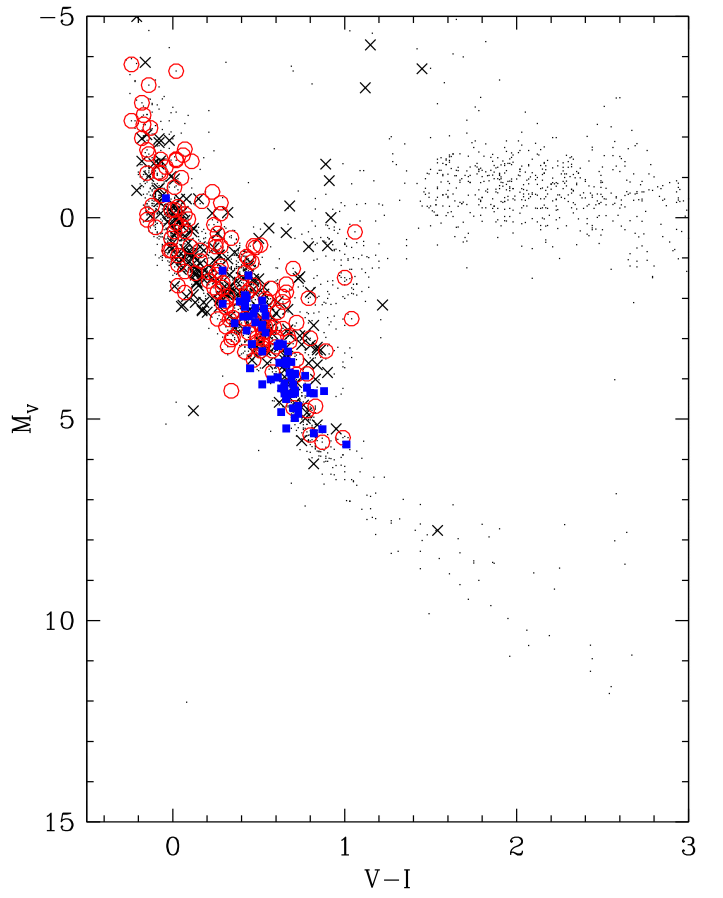}
	\end{center}
	\caption{\label{fig:HReb} HR Diagram with eclipsing binaries. The data are taken from the Annex "Periodic variables" in the Hipparcos catalogue. Crosses represent EA type binaries, open circles EB, filled squares EW.}
\end{figure}

The location of eclipsing binaries in the HR diagram is presented in figure~\ref{fig:HReb}. The three classical types of eclipsing binaries are presented: EA Algol type, EB $\beta$ Lyr type, and EW W UMa type. Note that other classifications of eclipsing binaries, probably more adequate to describe the physical properties of those systems, are available in the literature (e.g. \cite{pojmanski2002}, \cite{sarroetal2006}). It is interesting to note that EW types are confined in a smaller region in the HR diagram than EA and EB types.




\begin{figure}
	\begin{center}
		\includegraphics[width=120mm]{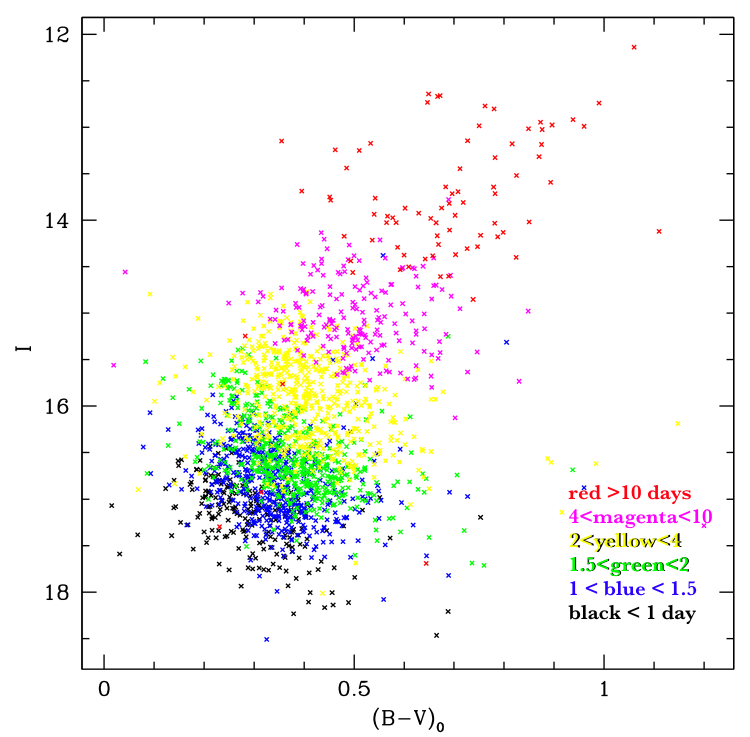}
	\end{center}
	\caption{\label{fig:HRcep} HR Diagrams with SMC OGLE-II cepheids.}
\end{figure}

We can have a closer look at the Cepheids in the SMC \cite{udalskietal1999}. OGLE-II provided a list of 2167 cepheids, that we can plot in a HR diagram. This is shown in figure~\ref{fig:HRcep}, where specific colours have been assigned for given ranges of pulsation periods. The periods are very clearly distributed, the longer ones being at higher luminosities and redder colours. This is expected from the period luminosity relation.

\begin{figure}
	\begin{center}
		\includegraphics[width=160mm]{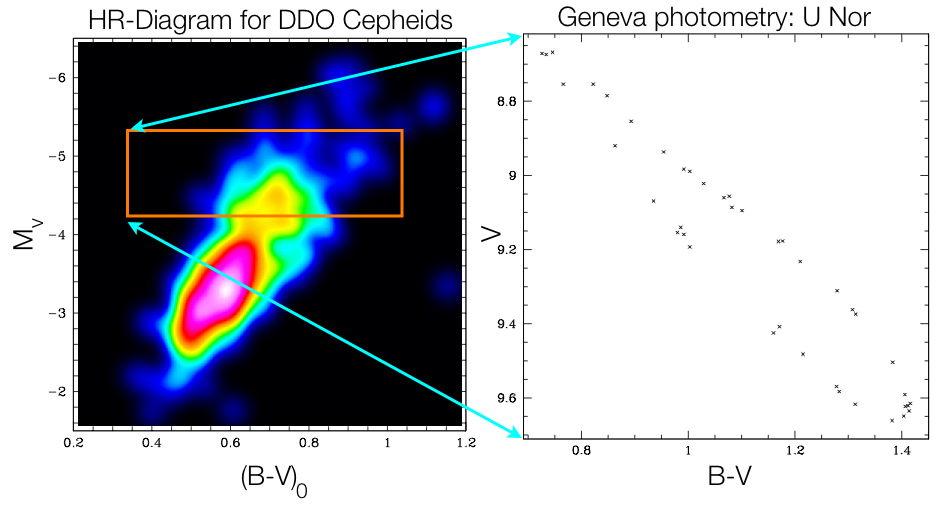}
	\end{center}
	\caption{\label{fig:HRmotion} Motion of the position of the cepheid U Nor in the HR diagram.}
\end{figure}

The motion of a cepheid in the HR diagram is presented in figure~\ref{fig:HRmotion}.

\section{Other observational diagrams}
\label{Sect:other diagrams}

Many diagrams other than the HR diagram presented in the previous section can be constructed based on quantities that can be readily derived from observational data.
We mention in this section some of them.
The list is not exhaustive, but gives a flavor of the importance of those diagrams in analyzing the properties of variable stars.

\subsection{Colour-Colour diagrams}

\begin{figure}
	\begin{center}
		\includegraphics[width=160mm]{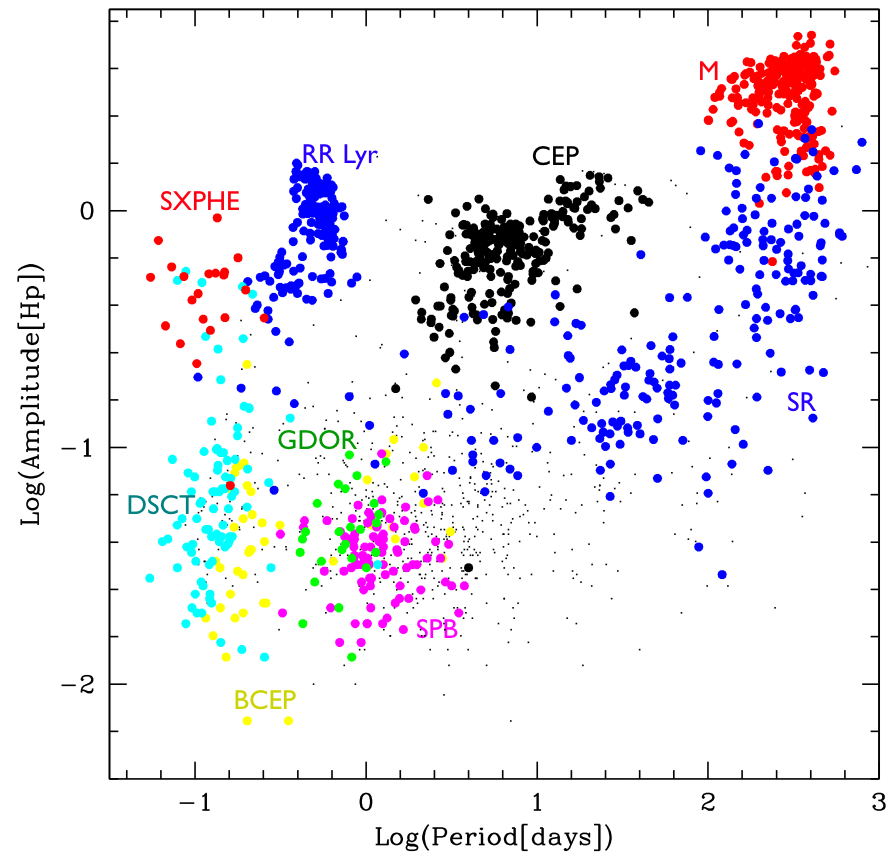}
	\end{center}
	\caption{\label{fig:P-A} Period-Amplitude diagram (logarithmic scale) for Hipparcos variable stars.}
\end{figure}

When the absolute magnitude cannot be obtained, we can still work on a colour-colour diagram. 
Such diagrams, however, are more degenerate than the classical HR diagram. There is for example an overlap between the region covered by the main sequence stars and the region covered by giant stars.

Nice examples of such colour-colour diagrams are given in \cite{sesaretal2007} for halo stars observed by the SDSS. In these diagrams, the authors provide information on various variability properties in the third dimension.

\subsection{Period-Amplitude diagrams}
In addition to the colours and luminosities, there are two other quantities that can easily be derived from the observation of variable stars.
Those are the period(s) and the amplitude(s) of variability measured from the light curves in one or several photometric bands.
These quantities are intrinsic properties of the star and are independent of the distance.
The amplitude of variability may, however, vary a lot depending on the photometric band considered.
The phase of the periodic signal may also vary with the photometric band.


The Period-Amplitude diagram constructed from the Hipparcos data of variable stars is shown in figure~\ref{fig:P-A}.
Expressed in logarithms of the periods and amplitudes, the diagram shows very clearly distinct regions of clumped data that correspond to different types of variable stars.
The combined usage of period and amplitude of variability is thus an efficient way to separate variability classes.
There are, however, regions that contain variable stars of different classes.
The region covered by $\beta$ Cep stars, for example, overlaps either with the region of slowly pulsating B stars, or with that of $\delta$ Scuti stars.
The region around $P=0.2$ and $A=0.3$ ($\log P=-0.7$ and $\log A=-0.5$) is another example of a region mixing several types of variable stars, gathering in this case SX Phoenicis, RR Lyrae and $\delta$ Scuti stars.

The diagram shown in figure~\ref{fig:P-A} is based on the variability type assignments listed in the "Catalogue of Variable Stars" of the Hipparcos mission at the exception of the $\gamma$ Dor stars, which were extracted from Simbad.

It must be noted that there has been no specific work devoted to the global systematic classification of variability types in the Hipparcos catalogue, at the exception for the distinction of eclipsing binaries (because of the doubling of the period), the RV Tauri stars (for the same reason), and for the B stars. Moreover, some variability class assignments were made based on information about the star taken from the literature, i.e outside the Hipparcos data.
It would be interesting to reanalyze the Hipparcos data with new classification methods based on machine learning techniques.
Such methods are under development in the framework of the preparation of the data analysis pipeline foreseen for the Gaia mission.
Preliminary results applied to the Hipparcos data are published in \cite{willemsenetal2005}.



\begin{figure}
	\begin{center}
		\includegraphics[width=160mm]{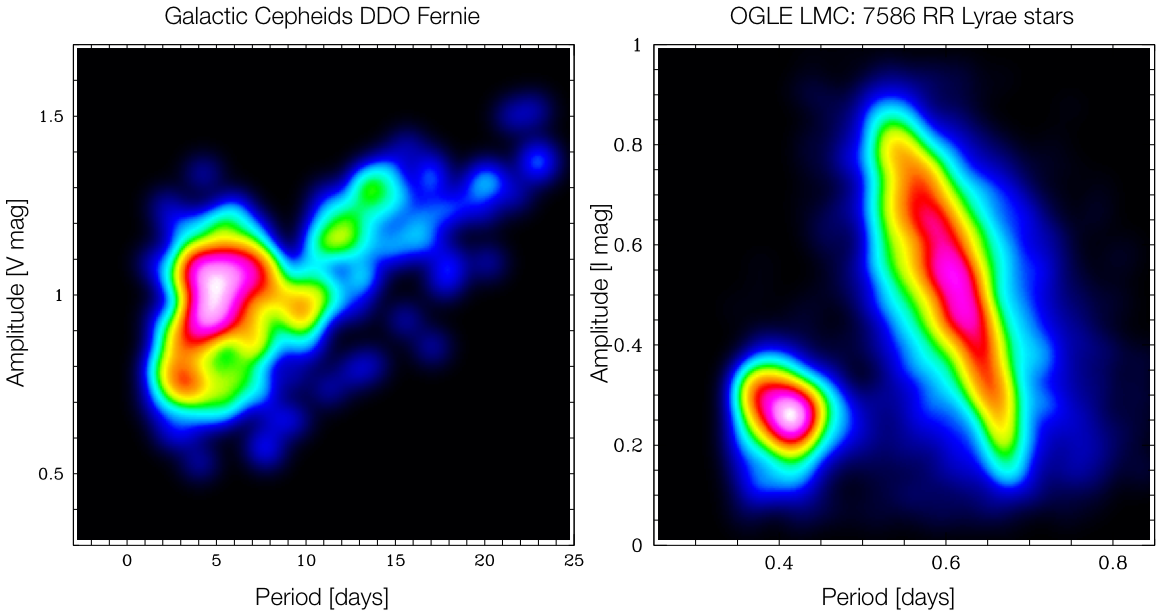}
	\end{center}
	\caption{\label{fig:probdistr} Probability distribution in the Period-Amplitude diagram for galactic cepheids (left figure, from the Fernie catalogue of the David Dunlap Observatory) and for RR Lyrae in the Large Magellanic Cloud (right figure, from the OGLE-II data base; the clump on the lower left identifies the RRc type while the elongated branch is the RRab type). }
\end{figure}

A probability distribution diagram can also be constructed for a given variability class, if the number of stars pertaining to the class is high enough.
Such a probability distribution in the Period-Amplitude diagram is shown in figure~\ref{fig:probdistr} for galactic cepheids (left figure) and for RR Lyrae in the Large Magellanic Cloud (right figure).
We see that the large surveys currently available allow now to obtain a statistical description of given populations of stars that is significant.
Caution must be taken, however, with the observational biases introduced, for example by the type of population observed and/or by the light curve sampling.
They can lead to biases in the distributions that can render their interpretation tricky.
Taking this into consideration, we can expect interesting and challenging results to be obtained in the near future from the statistical analysis of the data collected by large surveys, especially when comparing stellar populations differing in well defined parameters such as the metallicity.

Finally, we have to mention that, similarly to what may happen in the HR diagram for variable stars (see end of Sect.~\ref{Sect:HR}), a point representative of a given star may move in the Period-Amplitude diagram, depending on the time at which the amplitude and/or period are derived.
The amplitude of variability of an RR Lyrae may change in time due to the Blazhko effect.
Despite being known since almost a century, this phenomenon of amplitude and/or phase modulation (on a time scale five to ten times longer than the RR Lyrae period) still remains a puzzle for theoreticians.

\subsection{O-C diagrams}

The period of a variable star may also change with time, for many different reasons.
A usual way to analyse the period variability is to plot as a function of time the difference between the observed period and the one  expected in case of no period evolution.
Such diagrams are called O-C (Observed-Calculated) diagrams.
The slow evolution of periods observed in Cepheids, for example, has been analyzed in this way by \cite{pietrukowicz2003}, by comparing the data of OGLE and ASAS taken during the years 1997-2000 and Harvard data taken between 1910-1950.
We refer to the book of Sterken \cite{sterken2005} which gathers the proceedings of a conference entirely devoted to the analysis of periods using O-C diagrams.
Such period changes may lead to 'motions' in the Period-Amplitude diagram.



\subsection{Period-radius diagrams}

A rather recent breakthrough has been brought thanks to interferometry techniques.
The angular diameters and even their change caused by pulsation have been directly measured with interferometry. This allows to construct a new observational diagram, which is the period-radius diagram, see \cite{kervella2006} . Furthermore, the period luminosity relation can be derived avoiding the determination of the temperature.

\section{Some recent progress across the HR diagram}
\label{Sect:progress}

The field of variable stars and astero-seismology has been very active during the last decade, and substantial progresses have been achieved for several classes of variables.
In this section, we highlight some of them that cover different regions in the HR diagram.
The choice of variability classes is necessarily arbitrary, and by no means exhaustive. For example we do not mention $\alpha$ Cyg stars, Luminous Blue Variables (LBVs), or white dwarfs.
The purpose is to give a flavour of the on-going research activity.
We also note that some recent results found in the literature are still matter of debate, which after all is indicative of an actively evolving field.

We start with some types of variables found on the main sequence (Sect. 6.1-6.3) and consider red giants in Sect. 6.4. Some late phases of evolution are addressed in Sect. 6.5 to 6.7.

\subsection{Solar-like stars}

Evidence of solar-like oscillations was claimed by \cite{kjeldsenetal1995} in $\eta$ Boo, but the detection has been controversial.
The first strong identification of individual frequencies from such solar-like oscillations in a star other than the Sun was reported by \cite{bouchyetal2001} in $\alpha$ Cen A. This gave a real kick to the field of asteroseimology. The clear detection was made possible thanks to the improvement brought, within the planet search program, to the CORALIE instrument in measuring radial velocities. For $\alpha$ Cen A, amplitudes of 35 cm/s could be measured. Since then, the detection of solar-like oscillations has been reported for several other stars. We refer to the talk of Eggenberger in this Volume.

\subsection{Main sequence B-type stars}

Three types of variables are known among the mains sequence B stars: the $\beta$ Cephei stars, the Slowly Pulsating B (SPB) stars, and the Be stars.
The variability observed in $\beta$ Cephei and SPB stars is due to pulsation triggered by the $\kappa$ mechanism on the iron-group elements.
Main sequence B stars displaying the pulsational characteristics of both $\beta$ Cephei and SPB variables are also known.
Be stars, on the other hand, are defined by the presence of emission line(s) in their spectra arising from a circumstellar disk, and are characterized by irregular light curve variability on time scales from weeks to years related to the disk. But since the Be phenomenon is transient, those stars can also exhibit periodic variability due to the same $\kappa$ mechanism observed in other $\beta$ Cephei and SPB stars. Be stars are fast rotators.

The study of $\beta$ Cephei stars has witnessed great progress during the last decade thanks to sustained efforts to conduct multisite observations.
The astero-seismological potential of $\beta$ Cephei stars was re-emphasized when \cite{aertsetal2003} reported to be able to probe the inner regions of the star HD~129929 based on the analysis of photometric data gathered during 21 years.
They claimed to provide evidence for the occurrence of core convective overshooting, with an overshooting parameter between 5 and 15\% of the pressure scale height, as well as a hint for nonrigid rotation (\cite{dupretetal2004b}).
Multisite observation campaigns led thereafter to the publication of detailed astero-seismology analysis for $\nu$ Eridani (\cite{pamyatnykhetal2004},  \cite{ausseloosetal2004}) and several other $\beta$ Cephei stars (e.g. \cite{shobbrooketal2006}, \cite{mazumdaretal2006}). They led to the detection of several pulsating modes, and allowed to put constraints on the global stellar parameters as well as on the core overshooting parameter of the observed B stars.

The number of pulsating main sequence B stars is expected to decrease in stellar populations of decreasing metallicities, which seems confirmed by the observation of the number of pulsating B stars in several young open clusters of different metallicities \cite{pigulskietal2002B}.
This is due to the fact that the efficiency of the $\kappa$ mechanism on iron-group elements of course depends on the abundances of those elements.
Models predict the vanish of the $\beta$ Cephei instability strip at $Z<0.01$, while the SPB instability strip remains noticeable down to $Z \sim 0.006$.
Given the low metallicities of the Magellanic Clouds, they provide interesting laboratories to test these predictions.
Using the data from the large scale mircolensing surveys OGLE and MACHO, $\beta$ Cephei and SPB stars were discovered in the Large Magellanic Clouds (\cite{pigulskietal2002A}, \cite{kolaczkowskietal2004}).
But the metallicities of the stellar populations in the LMC span a large range, from $Z=0.004$ to 0.01 (\cite{maederetal1999}), making it difficult to check the number of pulsating B stars as a function of metallicity.
More convincing would be to check the presence (or absence) of variable B stars in the SMC.
Such a study has been recently undertaken by \cite{diagoetal2007}.
They found nine B stars showing short-period variability.
Surprisingly one of them is a $\beta$ Cephei variable, the eight remaining ones being SPB stars.
These authors also found three Be stars with pulsation periods shorter than 0.5 days, and suggest those to also be $\beta$ Cepheid stars.
If these result are confirmed, and if the metallicity of those proposed $\beta$ Cephei stars are indeed similar to the representative metallicitiy of the SMC, i.e. $Z \le 0.005$, then they provide a new challenge for stellar models.

\subsection{roAp stars}
 
Rapidly oscillating Ap stars are a subclass of chemically peculiar A stars which display photometric pulsations (non-radial) with periods between 5 and 20 minutes.
The small amplitudes of their variability, typically less than a few millimagnitudes, partly explain their relatively late discovery \cite{kurtz1982}.

Recent important progress was achieved in the study of those stars by analyzing the temporal variability of high-resolution spectral lines of different chemical elements.
Those lines are formed at different depth in the atmosphere, depending on the chemical element.
The pulsational characteristics result from the short vertical length of pulsation waves combined to the chemical stratification characteristic of the atmosphere of chemically peculiar stars.
Therefore, the analysis of the spectral lines variability provides information on the pulsation properties at different atmospheric depths, allowing to perform tomography of the atmosphere of those stars.

We refer to the reviews by \cite{sachkovetal2005} and \cite{kochukhov2007} for further details on the observation of pulsations in roAp stars.

\subsection{Red giants}

Several classes of variables are known since long among red giants. Initially, those were classified as Miras, semi-regular variables, and irregulars. The term Long Period Variable is also used. Because of the high amplitude variations of Miras, the variability of those giants could be studied since very early. A period luminosity relation (PL; see review by \cite{feast2004}) was obtained at the beginning of the twentieth century \cite{gerasimovic1928}. Since then and until about a decade ago, progress on the observational ground has led to a tightened relation, to the discovery of a second PL relation for semi-regular variables, parallel to that of Miras but shifted to shorter periods (\cite{woodetal1996}), and to the better knowledge that not only the variability amplitudes increase as the surface temperature of the red giant decreases (\cite{grenon1993}, \cite{eyeretal1994}) but also that {\it all} red giants with spectral types later than early K are variable, with a minimum variability level that increases with increasing spectral type (\cite{jorissenetal1997}, \cite{eyeretal1997}). 

The availability of large surveys in the last decade, such as MACHO and OGLE, opened a new era in the observational study of red giant variables. It is remarkable to see how the statistical analysis of tens of thousands of giants in the Magellanic Clouds led to the identification, so far, of fourteen distinct relations in the PL plane (\cite{soszynskietal2007} and references therein). The richness of information in those diagrams is such that it allows to split the near infrared PL sequences in the period-optical Wesenheit index plane into two separate ridges that seem to correspond to the spectral division of O-rich and C-rich AGB stars (\cite{soszynskietal2007}). Small amplitude variability has been highlighted among the OGLE red giants, baptised OGLE Small Amplitude Red Giants (OSARGs; \cite{wrayetal}), the nature of which is still unclear. The OGLE observations of the Magellanic Clouds also provide input to study Long Secondary Period variables (\cite{soszynskietal2007}).

On the theoretical ground, the physical origin of red giant variability remains  a subject of active research. Any quantitative analysis of variability in those stars requires a time-dependent description of convection, the coupling of which with pulsation is an important ingredient in providing reliable predictions. We discuss this point, and its results for the understanding of red giant variables, in section \ref{Sect:stellar evolution}.

%
%
%
%

\subsection{Pulsating sdB stars}

Pulsating B sub-dwarfs have been discovered in the last decade (\cite{kilkennyetal1997}; see \cite{kilkenny2007} for an observational review), and have since become a nice example of an astero-seismological field where great progress has been achieved in a short time. This new field has benefited from observational efforts that include long multi-site campaigns, both photometric and spectroscopic, as well as from progresses on the side of stellar modelling. A complete astero-seismological description of a dozen of sdB stars, leading to the precise estimation of the stellar properties (surface gravity, surface temperature, stellar mass, envelope mass, radius, luminosity, rotational period, ...), are available to date. We refer to the excellent review of \cite{fontaineetal2006}, in which an example of an end-to-end astero-seismological exercise can also be found.

There are about 70 pulsating sdB stars known so far, comprising two groups. One group consists in rapidly pulsating sdB stars, on time scales of few minutes, commonly known as EC~14026 variables. They are pulsating in p-modes. The second group of pulsating sdB stars, the so called \textit{Betsy}  variables, have periods of about 0.5 to 2.5 hours. They are g-mode pulsators.

The two groups define distinct regions in the $\log g - T\mathrm{_{eff}}$ plane. The EC~14026 stars occupy the high temperature side at $T\mathrm{_{eff}} \simgr 30,000$~K, while the \textit{Betsy}  stars are found below this temperature. Remarkably, two stars have been discovered at the boundary separating the two regions, that exhibit both p-mode and g-mode pulsations (\cite{schuhetal2006}, \cite{baranetal2005}, \cite{oreiroetal2005}).

The sdB stars are compact helium stars with masses around 0.5~M$_\odot$, that lie at the high temperature extension of the horizontal branch in the HR diagram. They are therefore also called Extreme Horizontal Branch stars. While the evolutionary scenario that leads to their formation is not yet known, the options being single star or binary evolutions, their structure is rather well defined. They are burning He in their core, but have only a thin H-rich envelope above it. Therefore, they will not be able to expand as an AGB after core He exhaustion, but will directly move left in the HR diagram to the white dwarf stage.

The pulsations observed in those stars originate from the $\kappa$ mechanism on iron group elements. For the mechanism to become efficient, however, diffusion processes must act in the envelope to increase the relative abundance of those heavy elements in their partial ionization zone. This is indeed predicted by evolutionary models of those stars. The existence of pulsating sdB stars was actually predicted at the same time than, but independently of, the discovery of EC~14026-2647, the first prototype of EC~14026 stars \cite{charpinetetal1996}.

It is the hope of the experts in the field to reach the point where enough statistics are reached in the properties of sdB stars to provide a stronger clue on their evolutionary origin.

\subsection{Post-AGB stars}

The region in the HR diagram occupied by post-AGB stars comprises several types of variable stars.
They include both H-rich stars such as RV~Tau stars, and H-depleted stars such as R CrB stars at the cool side of the HR diagram or PV Tel and extreme He stars at the hot side of that diagram (the case of GW Vir stars is presented separately in the next section).
Note that the distinction between RV~Tau and population II Cepheids is not always evident based solely on their pulsational characteristics (see the review of \cite{wallerstein2002}, for example).
From an observation point of view, most RV~Tau stars show infrared excesses, in contrast to Pop. II Cepheids, which is consistent with RV~Tau stars being post-AGB stars.

There has not been fundamental advances on the study of those variables in the last decade and
the field remains to be explored more deeply in future, most probably with the help of large scale surveys.
Some points are however worth mentioning.
RV~Tau stars are post AGB stars that become pulsationally unstable as they cross the classical instability strip during their motion to the hot side of the HR diagram.
In a recent study, \cite{kissetal2007} have been able to convincingly illustrate this fact by analyzing the pulsation properties of 17 post-AGB stars from the ASAS and the Northern Sky Variability Survey (NSVS).
The stars observed within the instability strip have a well defined periodicity, while stars lying at either edge of the strip are multiply periodic or possibly irregular.

H-deficient post-AGB stars are also expected to be studied in more detail during the coming years. The majority of them are thought to result from a last thermal instability that occurs in their helium-burning shell, either when they reach the hot side of the HR diagram or when they are already descending the white dwarf track. This late thermal pulse, as it is called, expands the thin envelope and turns the hot post-AGB star back to a red giant. Concomitantly, nucleosynthesis products of both the former hydrogen and helium burning shells are mixed into the envelope, resulting in a H-deficient envelope. While this scenario can explain the presence of most of the H-deficient stars, a fraction of them are almost exclusively dominated by pure helium. The formation of those extremely hot helium-rich stars may require a scenario involving the merge of two white dwarfs. They would also be the progeny of RCrB stars. We refer to the short conference article of \cite{rauchetal2006} for further references on this subject.

Finally, let us mention that the stellar evolutionary time scales of those post-AGB stars are relatively short. They could lead to secular evolution that could be detectable on human time scales (e.g. \cite{mowlavi1998}), such as witnessed by the well known stars FG Sagittae and the Sakurai object. The availability of surveys spanning several years may provide new opportunities to detect such objects. Needless to say, they would offer unique ways to test stellar evolution models.

\subsection{GW Vir stars}
There is a particular class of post-AGB stars that has attracted the attention of both observers and theoreticians in the past decade, some of those stars being among the best astero-seismically studied. Those are H-deficient post-AGB stars on their way to the white dwarf cooling sequence.
Those H-depleted atmosphere stars include the [WC] Wolf-Rayet central stars of planetary nebula, the PG~1159 stars that exhibit absorption lines of highly ionized He, C and O, and the DO He-dominated atmosphere white dwarfs. Those types of stars are evolutionary connected through the sequence [WC] $\rightarrow$ PG~1159 $\rightarrow$ DO. We refer to \cite{werneretal2006} for a review on those objects.

A good fraction of those H-deficient stars are pulsating. Various variability types are encountered in the literature, such as the GW~Vir stars, usually associated with PG~1159 variables, the PNNV, initially thought as the planetary nebula nucleus variables, and the DOV for the high surface gravity GW~Vir stars. As mentionned by \cite{quirionetal2007}, the terms PNNV and DOV are misleading and do not identify separate physical mechanisms for the pulsation. Therefore, we follow \cite{quirionetal2007} and use the name GW~Vir to qualify all variable H-deficient stars that lie in that part of the H-R diagram. They all have a common pulsation mechanism, which is the $\kappa$-mechanism associated with the partial ionization of carbon and oxygen (K-shell electrons).

It was \cite{starrfieldetal1979} who first suggested the existence of a new instability strip based on the $\kappa$ mechanism for partial ionization of C and O to explain the $\sim$539~s and $\sim$460~s pulsations detected in the newly discovered \cite{mcgrawetal1979} PG 1159-035 variable (=GW Vir in the naming of variable stars). PG~1159-035 would pulsate in low-order g-modes. The high effective temperature of this star (140,000~K) further suggested that the modes are non-radial (\cite{starrfieldetal1983}).

Since the discovery of the GW~Vir instability strip, controversial results were presented in the literature concerning the efficiency of C and O to induce the pulsations in the presence of He. The latter acts as a poison that smoothes the opacity gradients that would otherwise be obtained by the partial ionization of C and O (\cite{stanghellinietal1991}). Some authors confirm this difficulty (e.g. \cite{bradleyetal1996}, \cite{cox2003}), others obtain satisfactory results (\cite{saio1996}, \cite{gautschy1997}). The latest, comprehensive, analysis of GW Vir models seems to confirm the C-O ionization
origin of the pulsation, and shows that the extent of the instability domain in the log~$g$-$T_\mathrm{eff}$ plane is a strong function of the C and O abundance in their envelopes \cite{quirionetal2007}. The presence of pulsationally stable and unstable objects observed in that part of the H-R diagram is explained in terms of different chemical compositions in the envelope of those H-deficient stars.


\section{Some notes on stellar evolution models}
\label{Sect:stellar evolution}

The interpretation of the observed properties of variable and pulsating stars relies heavily on stellar models, whether to derive the global stellar parameters or to identify pulsation modes.
Our understanding of the different regions covered by variable stars in the HR diagram is also based on stellar evolution model predictions.
In this section, we highlight some development in this area during the last decade that has enabled us to better understand the variability properties across the HR diagram.
It is by far not exhaustive.
Among the important processes not covered here are rotation and chemical diffusion processes.

Convection is one of the physical processes that is difficult to be properly taken into account in stellar evolution models.
Being an essentially three dimensional problem, convection can only be included in one dimensional evolution models with a great deal of simplification.
While the classical parameterized mixing length theory is a convenient prescription to be numerically included in stellar evolution codes, it cannot describe satisfactorily the coupling between convection and pulsation.
Consequences of such shortcomings include, for example, the failure of 'standard' models to predict the correct location of the red edges of instability strips, the determination of which requires a non-adiabatic treatment of the interaction between convection and pulsation.
New time dependent models of convection have been developed in the last decade to overcome these shortcomings.
Successful results were obtained for RR Lyrae and Cepheids (\cite{xiongetal1998}), for $\delta$ Scuti stars (\cite{xiongetal2001}, \cite{dupretetal2004a}, \cite{dupretetal2005}), for $\gamma$ Dor stars (\cite{dupretetal2004a}, \cite{dupretetal2005}), for pre-main sequence stars (\cite{grigahceneetal2006}), and for red giants (\cite{xiongetal2007}).

The coupling of convection and pulsation is of particular importance for red giants, which are characterized by deep convective envelopes.
For example, \cite{xiongetal2007} convincingly claim to be able to predict, using appropriate time dependent models of convection, both the variability of red giants and the non-variability of stars lying, in the HR diagram, between the red giant branches and the classical instability strip of Cepheids/RR Lyrae.
They identify a 'red giant instability strip' in the lower temperature side of the HR diagram which goes along the sequences of the red giant branch and the asymptotic giant branch.
The excitation mechanism of that instability strip, they say, is related to the turbulent pressure of the convective region.
Their models further predict that red giants of low luminosity pulsate at high order overtones, the fundamental to second overtone being stable, while the instability moves to lower order modes when considering red giants of higher luminosity and lower effective temperature.
At the high luminosity end of the strip, red giants pulsate only in the fundamental and first overtone modes.
If confirmed, this scenario provides a nice explanation of many of the patterns observed in the variability zoo of red giants.
It also raises new issues. 
We refer to the paper of \cite{xiongetal2007} for further explanations and discussions.

The chemical abundance mixture is another parameter of stellar evolution that impacts model predictions.
For a given metallicity, the relative abundance of elements heavier than helium is usually scaled to that of the Sun, whose abundance was inferred, until recently, using 1D models of the solar atmosphere.
However, new 3D hydrodynamical models have led to a major revision of those solar abundances, resulting in significantly lower carbon, nitrogen, oxygen and neon abundances (\cite{asplundetal2005a}, \cite{asplundetal2005b}).
The new solar metallicity is now $Z=0.0126$, almost a factor of two lower than previously adopted.
While these revised abundances bring the solar abundance in agreement with that measured in the solar neighborhood (\cite{asplundetal2005b}), it destroys the remarkable agreement that was hitherto found between the sound speed profile observed by inverting helio-seismology data and that predicted by models (\cite{bahcalletal2005}).
The adoption of a new solar abundance mixture asks for a revision of the iron mass fraction derived at other metallicities as well.
This may be particularly important for $\beta$ Cephei and SPB stars that pulsate due to the $\kappa$ mechanism acting on the partial ionization zones of the iron-group elements.
And, indeed, a revision of the instability strip for those pulsating stars seems necessary.
We refer to \cite{miglioetal2007} for further discussion on this point.

\section{Conclusions}
\label{Sect:conclusions}

There is a flood of data being and becoming available thanks to a large number of ground based and space projects.
In many cases, there is a spectacular improvement in the precision (ground based radial velocity measurements or space-based photometric measurements), and new observable quantities such as the direct measurement of stellar radii with interferometry become possible.
We witness many advances in the knowledge of pulsation, but many things still remain to be understood.
We could paraphrase the cosmologist Mark Kamionkovski and conclude with conviction: "Now is the time to be an asteroseismologist".

\subsection{Acknowledgments}
We would like to thank the organiser of the meeting, Laurent Gizon. Our thanks go also to Mih\'aly V\'aradi, who helped us in some searches from the literature, and to Jan Cuypers for contributing to table~\ref{tab:NumberOfVariables}. 





\end{document}